\documentclass{jfm}{\tiny }

\usepackage{graphicx,natbib,empheq,relsize}  
\allowdisplaybreaks[1]

\ifCUPmtlplainloaded \else
  \checkfont{eurm10}
  \iffontfound
    \IfFileExists{upmath.sty}
      {\typeout{^^JFound AMS Euler Roman fonts on the system,
                   using the 'upmath' package.^^J}%
       \usepackage{upmath}}
      {\typeout{^^JFound AMS Euler Roman fonts on the system, but you
                   dont seem to have the}%
       \typeout{'upmath' package installed. JFM.cls can take advantage
                 of these fonts,^^Jif you use 'upmath' package.^^J}%
      }
  \else
  \fi
\fi

\ifCUPmtlplainloaded \else
  \checkfont{msam10}
  \iffontfound
    \IfFileExists{amssymb.sty}
      {\typeout{^^JFound AMS Symbol fonts on the system, using the
                'amssymb' package.^^J}%
       \usepackage{amssymb}%
         \let\leq=\leqslant
         \let\geq=\geqslant
      }{}
  \fi
\fi

\ifCUPmtlplainloaded \else
  \IfFileExists{amsbsy.sty}
    {\typeout{^^JFound the 'amsbsy' package on the system, using it.^^J}%
     \usepackage{amsbsy}}
    {\providecommand\boldsymbol[1]{\mbox{\boldmath $##1$}}}
\fi

\providecommand\bcdot{\boldsymbol{\cdot}}
 
\usepackage{mathrsfs} 
\usepackage{multirow}
\usepackage{amsmath,mathtools}
\usepackage[usenames,dvipsnames]{xcolor} 

\newcommand{\tr}{\textrm{tr}\,}
\newcommand{\sym}{\textrm{sym}\,} 
 
\newcommand{\Pos}{{\textbf{Pos}}}

\newcommand{\bld}[1]{{\boldsymbol{ #1 }} } 
\DeclareMathAlphabet{\mathscrbf}{OMS}{mdugm}{b}{n} 
\newcommand\Wie{\mbox{\textit{Wi}}}  
\newcommand{\e}[1]{{\text{e}^{ #1 }} } 

\def\sldsh{\rule[0.2\baselineskip]{0.075in}{0.65pt}}
\def\sldot{\rule[0.2\baselineskip]{0.025in}{0.65pt}}
\def\bldot{\hspace{0.025in}}
\def\linesolids{\rule[0.2\baselineskip]{0.275in}{0.65pt}}
\def\linesolidsthick{\rule[0.2\baselineskip]{0.275in}{0.975pt}}
\def\linedashed{\sldsh\bldot\sldsh\bldot\sldsh} 
\def\linedshdot{\sldsh\bldot\sldot\bldot\sldsh\bldot\sldot}

\def\sldshthick{\rule[0.2\baselineskip]{0.075in}{1.5pt}}
 
\def\linesolidsthick{\rule[0.2\baselineskip]{0.275in}{1.5pt}}
\def\linedashedthick{\sldshthick\bldot\sldshthick\bldot\sldshthick}

\def\linesolidsCirc{{\color{NavyBlue!85} \rule[0.2\baselineskip]{0.1375in}{1.00pt}\hspace{-0.01in}\parbox{0.0825in}{$\vspace{-0.0075in}\mathlarger{\mathlarger{\mathlarger{\circ}}}$}\rule[0.2\baselineskip]{0.1375in}{1.00pt}}}
\def\linesolidsDiam{{\color{Red} \rule[0.2\baselineskip]{0.1375in}{1.00pt}\hspace{-0.01in}\parbox{0.0825in}{$\vspace{-0.0075in}\mathlarger{\mathlarger{\mathlarger{\diamond}}}$}\hspace{0.005in}\rule[0.2\baselineskip]{0.1375in}{1.00pt}}}
\def\linesolidsSqr{{\color{YellowOrange} \rule[0.2\baselineskip]{0.1375in}{1.00pt}\hspace{-0.01in}\parbox{0.0825in}{$\vspace{-0.0075in}\mathlarger{{{\square}}}$}\rule[0.2\baselineskip]{0.1375in}{1.00pt}}}
\def\linesolidsTri{{\color{LimeGreen} \rule[0.2\baselineskip]{0.1375in}{1.00pt}\hspace{-0.03in}\parbox{0.0825in}{$\vspace{-0.0075in}\mathlarger{{{\triangle}}}$}\hspace{0.005in}\rule[0.2\baselineskip]{0.1375in}{1.00pt}}}

 \newlength{\myMheight}
\newcommand{\rev}[1]{{#1}}

\def\tz#1{\textcolor{orange}{#1}}

\newcommand{\am}[1]{\langle #1 \rangle_{\scriptscriptstyle \sum}} 
\newcommand{\gm}[1]{\langle #1 \rangle_{\scriptscriptstyle \prod}} 
\newcommand{\lEm}[1]{\langle #1 \rangle_{\log}} 
\newcommand{\sqrtm}[1]{\langle #1 \rangle_{\sqrt{\,}}} 
\newcommand{\circm}[1]{\langle #1 \rangle_{\circ}}

\title[The mean conformation tensor in viscoelastic turbulence]{The mean conformation tensor in viscoelastic turbulence}

\author[Hameduddin \& Zaki]%
{Ismail Hameduddin\aff{1} 
	and Tamer  A. Zaki\aff{2}\corresp{\email{t.zaki@jhu.edu}} 
}

\affiliation{
  \aff{1} Department of Mathematics, University of British Columbia,  Vancouver, BC V6T1Z2, Canada 
\aff{2}	Dept. of Mechanical Engineering, The Johns Hopkins University,
	Baltimore, MD 21218, USA }

\pubyear{2010}
\volume{650}
\pagerange{119--126}
\date{\today; revised xx?; accepted ?. - To be entered by editorial office}

\begin{document}
  \maketitle 
\begin{abstract}
This work demonstrates that the popular arithmetic mean conformation tensor frequently used in the analysis of turbulent viscoelastic flows is not a good representative of the ensemble.
\rev{Alternative means based on recent developments in the literature 
are proposed, namely, the geometric and log-Euclidean means.
These means are mathematically consistent with the Riemannian structure of the manifold of positive-definite tensors, on which the conformation tensor lives, and have useful properties that make them attractive alternatives to the arithmetic mean.}
Using a turbulent FENE-P channel flow dataset, it is shown that these two alternatives are physically representative of the ensemble. 
\rev{By definition these means minimize the geodesic distance to realizations and exactly preserve the scalar geometric mean of the volume and of the principal stretches.}
The proposed geometric and log-Euclidean means have a clear physical interpretation and provide an attractive quantity for turbulence modelling.  
\end{abstract}

\section{Introduction}
Fluid turbulence is defined by irregular and chaotic `eddying' motions, whose characterization has been subjected to intense study over the last century. 
In practical applications, one usually separates out a statistically persistent state of the turbulence, and analyzes fluctuations with respect to this mean state. 
This simple, yet revolutionary, perspective 
originated in the kinetic theory of gases  
and was first introduced by Reynolds.  
Since then, the mean--fluctuation separation has been adopted as the staple approach to quantify turbulence.
The mean velocity is typically calculated using standard arithmetic averaging in some or all of the statistically homogeneous spatiotemporal directions and is commonly accepted to be a good first-order approximation of the actual instantaneous velocity. It is thus widely used to study the flow physics and also for turbulence modelling.

In viscoelastic flows, the velocity field is insufficient to describe the dynamics because it necessarily depends on the flow deformation history.
The latter is encoded in a second-order positive-definite tensor known as the conformation tensor.
In analogy with the approach adopted for the velocity field, arithmetic averaging 
is typically assumed to yield a representative conformation tensor.
Such an approach has the convenience of maintaining consistency with the averaging approach used for the velocity field. 
In addition, the resulting mean conformation tensor appears directly in the mean momentum equation for some common constitutive models.
As such, the arithmetic mean conformation tensor has often been adopted in the analysis and modelling of viscoelastic turbulence \citep{Housiadas2003,Masoudian2013,Lee2017}.

Here, we show that the arithmetic mean conformation is a poor representative of the typical conformation.
As a result, using it to infer the flow physics may lead to erroneous conclusions, and also to difficulties in turbulence modelling because it would limit the role played by physical intuition to derive closures and reduced-order models.
We propose alternative means based on recent results that rely on \rev{Riemannian geometries} natural to the set of positive-definite tensors, and use these means to examine drag-reduced channel flow.

\rev{For the majority of the present work,} we use a dataset of 60\% drag-reduced turbulent channel flow of a dilute polymer (FENE-P) solution obtained from direct numerical simulations (DNS) reported by \citet{Hameduddin2018a}. 
The constant mass flow rate channel flow simulation is periodic in the horizontal directions and the relevant parameters are given in table \ref{tab:SimParams}.
Detailed descriptions of the DNS are provided in \citet{Hameduddin2018a}.
\rev{Where data from additional simulations are reported, the setup is identical except for the value of the Weissenberg number which is varied in order to assess the influence of elasticity.}

In \S\ref{sec:arithmetic} we discuss the deficiency of the arithmetic mean. 
Alternatives are proposed  in \S\ref{sec:altmeans}, and evaluated in FENE-P turbulent channel flow in \S\ref{sec:dns}. Conclusions are offered in \S\ref{sec:conclusions}.

\begin{table}
\centering
    \begin{tabular}{ccccccccc}
      $\Rey$ & $\Rey_{\tau}$ & $\Wie$ & $L_{\max}$ & $\beta$ & 
      $L_x\times L_y \times L_z$ & $N_x\times N_y \times N_z$ & $\Delta t$ 
      & $\Delta_x^+ \times\Delta_y^+ \times\Delta_z^+$ \\
      4667 & 180 & 6.67 & 100 & 0.9 & $4\pi \times 2 \times 4\pi$ &
      $512 \times 400 \times 512$ &  $0.0025$ & $4.42\times[0.13,1.90]\times 4.42$ 
    \end{tabular} 
  
  \caption{Parameters of the simulation of viscoelastic turbulent channel flow: $\Rey$ is the Reynolds number, $\Rey_{\tau}$ is the frictional Reynolds number, $\Wie$ is the Weissenberg number, $L_{\max}$ is the maximum polymer extensibility, and $\beta$ is the viscosity ratio.
    The length scale is the channel half-height and velocity scale is the bulk flow speed. The friction velocity is approximated using the mean velocity gradient at the wall. The size of the domain in the direction $p$ is $L_p$, while $N_p$ is number of points. The spatial resolution in direction $p$ is $\Delta_p$.}
  \label{tab:SimParams}
\end{table}

\section{Problem of arithmetic mean}
\label{sec:arithmetic}

The utility of the arithmetic mean in representing an ensemble is not always guaranteed: 
the arithmetic mean of a set of rotation matrices is, in general, not a rotation matrix and thus cannot possibly be representative of a typical rotation. 
As will be shown, the arithmetic mean of the conformation tensor is not representative, even though it is still a positive-definite tensor.
\citet{Hameduddin2018a} found that the volume of the arithmetic mean conformation tensor is several orders of magnitude larger than the typical realization, and demonstrated that evaluating fluctuations about such a mean may not be meaningful. 
This volume `swelling' of the arithmetic mean has been found in other contexts as well \citep{Arsigny2007}.

One intuitive way to characterize the conformation tensor, $\mathsfbi{C}$, is by examining its eigenvalues, which represent the squared lengths along the principal axes of the deformation.
The eigenvalues are strictly positive, and expansions with respect to the thermodynamic equilibrium lead to eigenvalues $> 1$, while compressions lead to eigenvalues $< 1$.
Thus $A=\tr \sqrt{\mathsfbi{C}}$ and $V=\sqrt{ \det \mathsfbi{C}}$ are proportional to the average stretch and the volume, respectively.
A good proxy for the surface area of the ellipsoid represented by the principal axes and stretches of $\mathsfbi{C}$ is  $S=(1/2)[(\tr \sqrt{\mathsfbi{C}})^2 - \tr\mathsfbi{C}]$.

\begin{figure}
	\centering
	\includegraphics[scale=1.0]{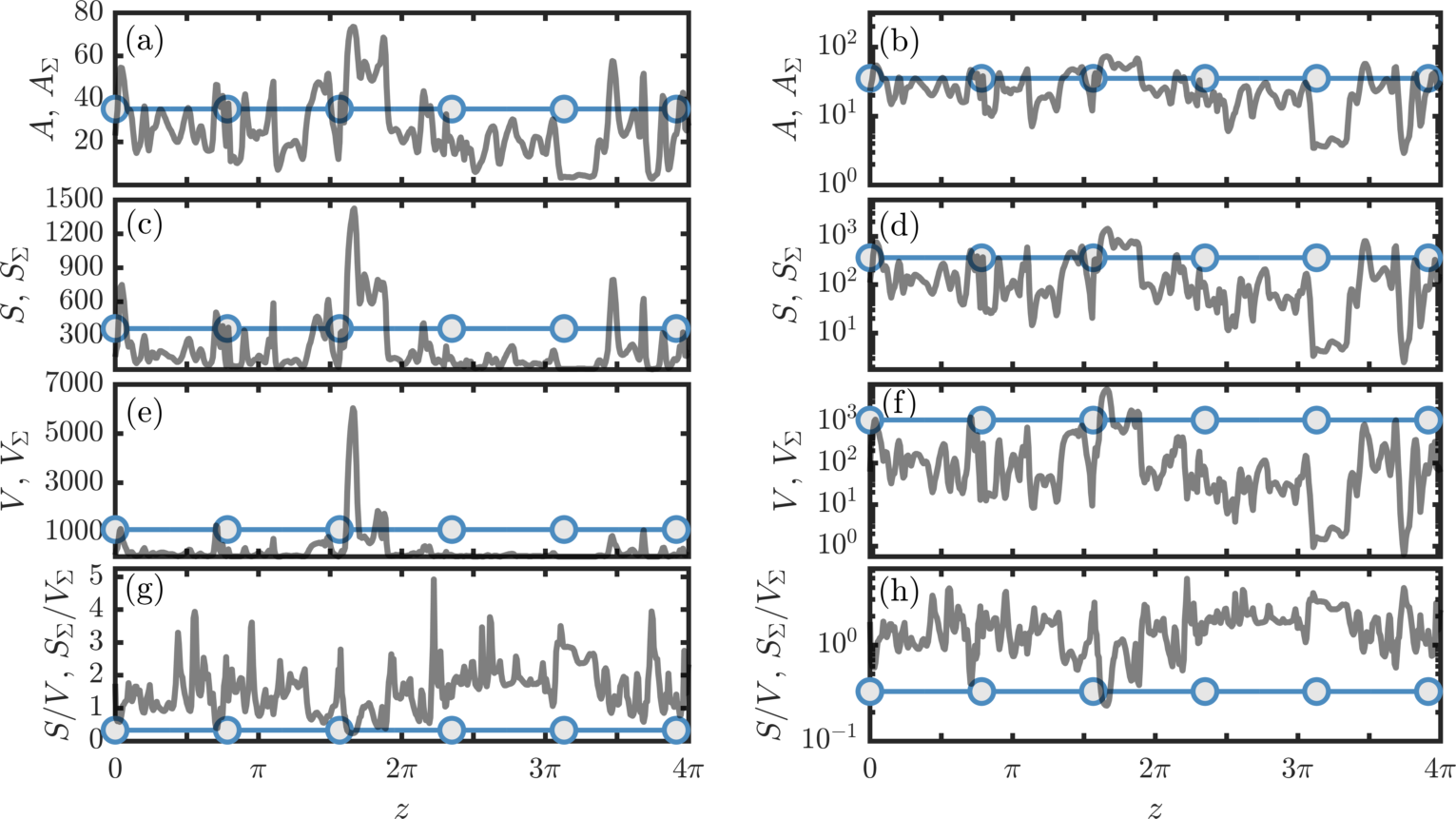}
	\caption{\rev{Characterization of conformation tensor: (a)--(b) sum of stretches, $A$ and $A_{\scriptscriptstyle \sum}$; (c)--(d) surface area, $S$ and $S_{\scriptscriptstyle \sum}$; (e)--(f) volume, $V$ and $V_{\scriptscriptstyle \sum}$; (g)--(h) surface area to volume ratio, $S/V$ and $S_{\scriptscriptstyle \sum}/V_{\scriptscriptstyle \sum}$. Fluctuating quantities along a $z$ traverse at $y^+=100$ (solid black line with 50\% transparency,{\color{gray}\linesolidsthick}) and quantities calculated using the arithmetic mean along the traverse (blue line with circle markers,
	\linesolidsCirc). The left column represents quantities on a linear scale and the right column represents the same quantities on a logarithmic scale.}}
	\label{fig:am_shape_lin_log}
\end{figure}

\rev{The quantities $A$, $V$, $S$, and $S/V$ in drag-reduced viscoelastic FENE-P turbulent channel flow are shown in figure \ref{fig:am_shape_lin_log}.
Also shown are corresponding quantities calculated from the arithmetic mean conformation tensor: $A_{\scriptscriptstyle \sum}=\tr \sqrt{\am{\mathsfbi{C}}}$, $V_{\scriptscriptstyle \sum}=  \sqrt{\det\am{\mathsfbi{C}}}$, $S_{\scriptscriptstyle \sum}= (1/2)[(\tr \sqrt{\am{\mathsfbi{C}}})^2 - \tr \am{\mathsfbi{C}}]$ and $S_{\scriptscriptstyle \sum}/V_{\scriptscriptstyle \sum}$, where $\am{\mathsfbi{C}}$ is the arithmetic mean of $\mathsfbi{C}$. }
The quantities are along a spanwise ($z$) traverse at wall-normal location $y^+=100$, where $+$ superscript indicates friction units, and an arbitrary streamwise ($x$) location.
The averaging 
is performed along the traverse.

The average stretch $A$ shown in figure \ref{fig:am_shape_lin_log}(a) appears to be captured well by the arithmetic mean conformation tensor. However, $S$ and $V$ shown in figures \ref{fig:am_shape_lin_log}(c,e) are not captured well, with the arithmetic mean having much larger surface area and volume. 
The ratio $S/V$, shown in figure \ref{fig:am_shape_lin_log}(g), is a proxy for inverse of sphericity since it is minimized when all the stretches are equal.
The sphericity associated with the arithmetic mean is strikingly larger than that of the typical conformation tensor along the traverse, demonstrating that the shape of the former is not representative and cannot be easily used to infer the physics of turbulent polymer deformation.

\begin{figure}
  \centering
  \includegraphics[scale=1.0]{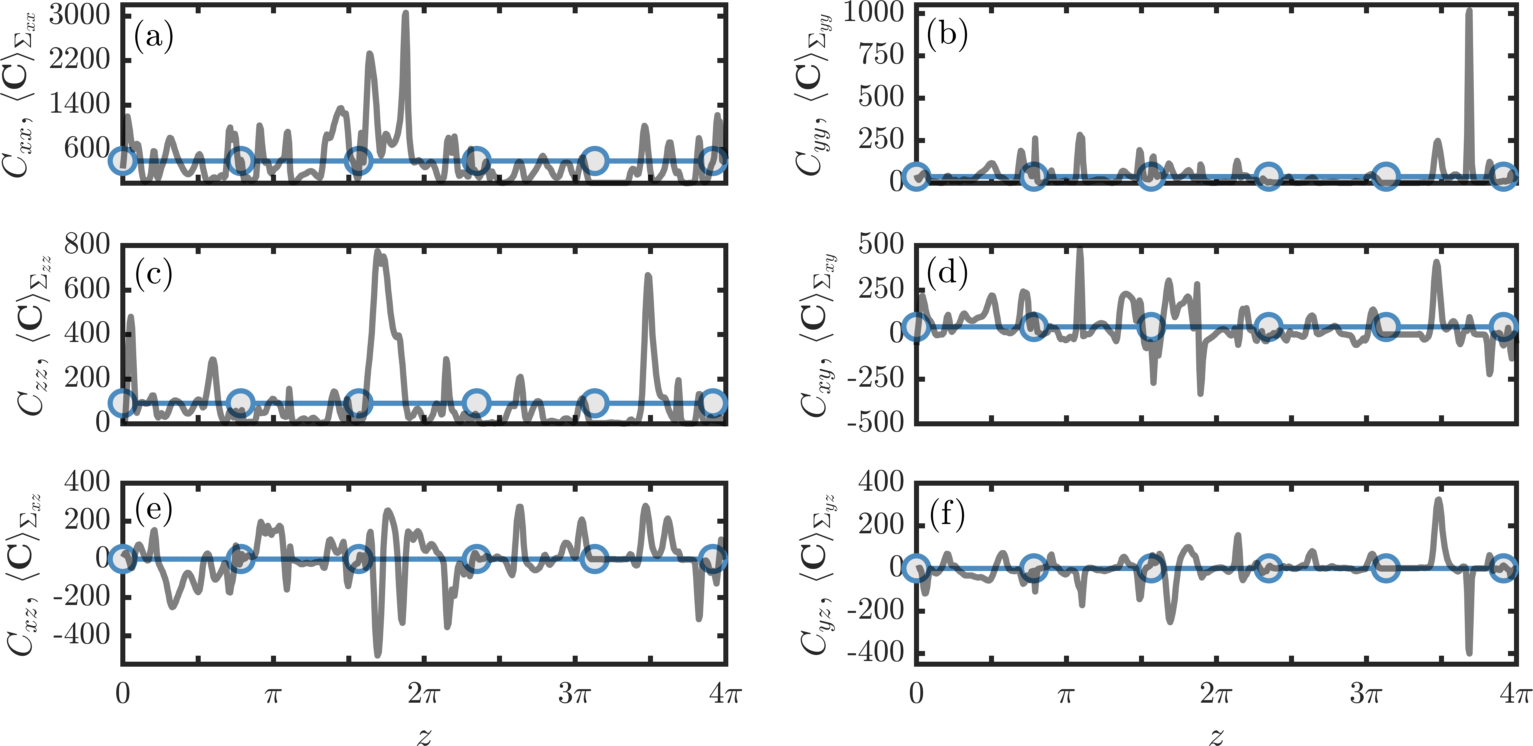}
  \caption{\rev{Fluctuating components of conformation tensor  along a $z$ traverse at $y^+=100$ (solid black lines with 50\% transparency,{\color{gray}\linesolidsthick}) compared to components of the arithmetic conformation tensor along the traverse (blue lines with circle markers,
    \linesolidsCirc): (a) $xx$, (b) $yy$, (c) $zz$, (d) $xy$, (e) $xz$ and (f) $yz$. }}
  \label{fig:C_am_trace}
\end{figure}


The eigenvalues of $\mathsfbi{C}$ are restricted to $(0,1)$ for compressions and $(1,\infty)$ for expansions and thus the latter will appear larger on a linear scale than a corresponding compression.
This behaviour reflects the logarithmic nature of the principal stretches of the conformation tensor and accordingly leads to the large excursions evidenced in the left column of figure \ref{fig:am_shape_lin_log}.
We have therefore plotted $A$, $V$, $S$, and $S/V$ on a logarithmic scale in the right column of figure \ref{fig:am_shape_lin_log}. 
On this scale, $V$, $S$, and $S/V$ are more ill-represented by the arithmetic mean conformation tensor, while $A$ appears to be now over-predicted.

\rev{The measures in figure \ref{fig:am_shape_lin_log} are holistic in the sense that they characterize the conformation tensor as a whole rather than individual components.
Figure \ref{fig:C_am_trace} shows the individual components of the conformation tensor and the arithmetic mean conformation tensor along the same $z$ traverse shown in figure \ref{fig:am_shape_lin_log}.
The normal components are dominated by the streamwise component except for brief intervals during which $\mathsfi{C}_{yy}$ and $\mathsfi{C}_{zz}$ are dramatically elevated.
The normal components are restricted to be greater than zero, similar to the principal stretches.
No such restriction is imposed on the cross-components, which also show a large excursions, both positive and negative.
As expected, the components of the arithmetic mean conformation tensor appear representative of the respective components of $\mathsfbi{C}$ when plotted on a linear scale.
However, as discussed earlier, the arithmetic mean conformation tensor as a whole is not representative of the ensemble, especially when the associated deformation is evaluated on a logarithmic scale, which is the natural scale when representing compressions and expansions.
The large excursions in the normal components noted earlier are considerably less severe on such a scale.
It is not clear how to evaluate the large excursions in the cross-components since these take on both positive and negative values.}

The mathematical reason for the ill-suitability of the arithmetic mean conformation tensor is rooted in the geometry of the set of positive-definite tensors, here referred to as $\Pos_3$, which is not Euclidean.
For instance, consider the straight line  
\begin{align}
\mathsfbi{X}(r) = (1-r)\mathsfbi{A} + r\mathsfbi{B} \qquad r \in \mathbb{R}, \qquad\mathsfbi{A},\mathsfbi{B}\in \Pos_3.  \label{EuclidPath}
\end{align} 
It is straightforward to show that for $|r|\gg 1$, $\mathsfbi{X}(r)$ loses positive-definiteness, i.e. exits the set $\Pos_3$, thereby violating Euclid's second postulate.
In fact, it can be shown that $\mathsfbi{X}(r)$ is guaranteed to be in $\Pos_3$ only for $ r \in [0,1]$.
As a result of the lack of Euclidean structure to $\Pos_3$, Euclidean geometric notions, e.g. distances, do not apply. 
\rev{The lack of Euclidean structure on $\Pos_3$ is a direct reflection of the fact that compressions and expansions lead to eigenvalues of $\mathsfbi{C}$ in $(0,1)$ and $(1,\infty)$, respectively, which is partially the reason for ill-suitability of arithmetic mean: averaging compressions and expansions in the usual sense would bias the mean towards expansions.
In addition, the rotation of the principal axes of $\mathsfbi{C}$ introduces new complications because deformations are then not averaged along unique directions anymore as assumed above. 
}

To see more directly why the standard arithmetic mean is problematic on $\Pos_3$, it is instructive to consider an alternative characterization of the arithmetic mean.
Without loss of generality, assume the conformation tensor $\mathsfbi{C}(x)$ is to be averaged over $x \in \Omega \subseteq \mathbb{R}$.
\rev{The Fr\'{e}chet mean is given by
\begin{align}
{\arg\min}_{\mathsfbi{W} \in \Pos_3} \int_{\Omega} \textrm{dist}(\mathsfbi{C}(x),\mathsfbi{W}) \, \text{d}x \label{FrechetDefn}
\end{align}
where $\textrm{dist}(\mathsfbi{C}(x),\mathsfbi{W})$ is a given function measuring distance between $\mathsfbi{C}(x)$ and $\mathsfbi{W}$.
Geometrically, the Fr\'{e}chet mean is the centroid\textemdash in terms of the distance function chosen \textemdash of the set of conformation tensors represented by $\mathsfbi{C}(x)$.
If the distance function is chosen to be the Frobenius norm $\| \cdot \|_F$, the Fr\'{e}chet mean reduces to the standard arithmetic mean, i.e.
}
\begin{align}
\am{\mathsfbi{C}} =
{\arg\min}_{\mathsfbi{W} \in \Pos_3} \int_{\Omega} \|\mathsfbi{C}(x) - \mathsfbi{W}\|_F^2 \, \text{d}x. \label{amFrechet}
\end{align}   
The problem with such a characterization in the form \eqref{amFrechet} is that $\mathsfbi{C}(x) - \mathsfbi{W}$ assumes that the Euclidean straight line is the relevant `shortest path' between $\mathsfbi{C}(x)$ and $\mathsfbi{W}$.
However, as described earlier $\Pos_3$ is not a Euclidean space and therefore Euclidean paths are not relevant.

\rev{Specifically, the shortest path joining two points cannot be indefinitely extended and thus violates our intuitive understanding of how well-behaved shortest paths should behave: they should be indefinitely extendible. Thus, the Euclidean distance (the Frobenius norm) associated with the shortest path is not appropriate.
It is instructive here to consider the scalar case: if $s$ represents the length of a uniaxial rod undergoing deformation, then because the rod cannot be compressed to zero length, $s=0$ is unattainable and is thus infinitely far from every other value $s$ can take. The standard Euclidean distance does not treat $s=0$ as such\textemdash it is possible to go from $0<s<\infty$ to zero length by finitely extending a Euclidean path\textemdash and accordingly the Fr\'{e}chet mean with such a distance would not yield a representative mean.
We can expect the provision of appropriately formulated infinitely extendible shortest paths to overcome this problem as then $s=0$ would have to be infinitely far from every other value of $s$. }

Since the underlying problem with the arithmetic mean is related to the geometric structure on $\Pos_3$, the challenge is to formulate a mean conformation tensor based on the geometry natural to $\Pos_3$.
This challenge has been addressed in the mathematical literature in various forms \citep{Ando2004,Moakher2005,Arsigny2006}.
In \S \ref{sec:altmeans}, we present the principal results that provide better alternatives to the arithmetic mean.

\section{Alternative means}
 
\label{sec:altmeans}
 
The set $\Pos_3$ can be endowed with a powerful geometric structure; it is a simply connected, geodesically complete Riemannian manifold \citep{Lang2001}.
In this geometry, straight lines are replaced by geodesic curves, which are guaranteed to be unique, smooth and indefinitely extendible.
\rev{The guarantee on indefinite extension of geodesic curves is provided by the geodesic completeness property and is a key component of constructing a geometry that provides an intuitive analog to Euclidean geometry and one which will then lead to reasonable definitions of the mean.}
The Frobenius norm, which \rev{measures} Euclidean distances, is replaced by the geodesic distance.
The geodesic distance  between $\mathsfbi{A},\mathsfbi{B}\in \Pos_3$ \rev{associated with the Rao--Fisher metric on $\Pos_3$} is 
\begin{align}
d(\mathsfbi{A},\mathsfbi{B}) = 
\sqrt{\tr \log^2(\mathsfbi{A}^{-\frac{1}{2}} \bcdot
  \mathsfbi{B} \bcdot 
  \mathsfbi{A}^{-\frac{1}{2}})}. \label{geodesicDefn}
\end{align}
\rev{The set $\Pos_3$ forms a complete metric space under the distance function \eqref{geodesicDefn} as a consequence of the well-behaved geometric structure.}

The distance function in \eqref{geodesicDefn} exhibits several features \rev{that are analogous to the notion of distance in Euclidean space}. Analogous to the translation invariance \rev{of distance in Euclidean space}, the geodesic distance is invariant under the action of the general linear group:
$d(\mathsfbi{A},\mathsfbi{B})  = 
d(\mathsfbi{Y} \bcdot\mathsfbi{A}\bcdot \mathsfbi{Y}^{\mathsf{T}} ,\mathsfbi{Y} \bcdot \mathsfbi{B} \bcdot \mathsfbi{Y}^{\mathsf{T}} )$
for any invertible $\mathsfbi{Y}$.
\rev{The action of the general linear group has a physical interpretation as the application of a deformation since the conformation tensor can be viewed as a left Cauchy-Green tensor with deformation gradient $\mathsfbi{C}^{1/2}$. Then $\mathsfbi{Y} \bcdot\mathsfbi{C}\bcdot \mathsfbi{Y}^{\mathsf{T}} = (\mathsfbi{Y} \bcdot\mathsfbi{C}^{1/2})\bcdot(\mathsfbi{Y} \bcdot\mathsfbi{C}^{1/2})^{\mathsf{T}}$ implies that $\mathsfbi{Y} \bcdot\mathsfbi{C}^{1/2}$ is the new total deformation gradient.
Then invariance under this action means that the distance between two tensors remains the same if both are simultaneously deformed in the same way, which is a physically intuitive notion.
\cite{Hameduddin2018a} used this interpretation to formulate an alternative to the Reynolds decomposition that is appropriate for conformation tensors, while \cite{Hameduddin2019a} used it to derive perturbative expansions to the conformation tensor.
}

\rev{Another important property of the distance function is invariance under inversion:
$d(\mathsfbi{A},\mathsfbi{B})  = 
d(\mathsfbi{A}^{-1},\mathsfbi{B}^{-1})$.
This property  is especially attractive because it implies that expansions and compressions are treated on an equal footing, unlike with the Euclidean distance function. It is also analogous to the invariance of distance in Euclidean space under reversal of direction.
}

\rev{Although the particular Riemannian structure described above is well-known in the mathematical literature, it has not been exploited in studies of viscoelastic flows.  In addition, it is not the only one possible for the set of positive-definite tensors, e.g. \citet{Hiai2009} introduce a family of Riemannian metrics defined by a kernel function and of which the present metric is a member. 
However, the present Riemannian metric brings together the above described distinct features that make the geometry analogous to the that used in Euclidean spaces and also physically intuitive: 
it is geodesically complete, it is invariant under the action of the general linear group and it is invariant under inversions.
Another interesting geodesically complete Riemannian metric to be introduced later, the log-Euclidean metric, is invariant under inversions but is not invariant under the action of the general linear group.
}

The distance function \eqref{geodesicDefn} allows us to formulate an alternative to \eqref{amFrechet} by replacing the Euclidean distance with the geodesic one; the geometric mean is defined as
\begin{align}
\gm{\mathsfbi{C}} \equiv
{\arg\min}_{\mathsfbi{W}\in\Pos_3} \int_{\Omega} d^2(\mathsfbi{C}(x), \mathsfbi{W}) \, \text{d}x. \label{geomMeanDefn}
\end{align} 
The geometric mean defined in \eqref{geomMeanDefn} was proposed by \citet{Moakher2005} and can be shown to be a generalization of the scalar counterpart $(\prod_{i=1}^M a_i)^{1/M}$ for a set of $M$ numbers $\{a_i\}_{i=1}^M$.
\rev{By the inversion invariance of $d(\cdot,\cdot)$, the geometric mean of $\mathsfbi{C}^{-1}$ is precisely $\gm{\mathsfbi{C}}^{-1}$.
This notable result shows that expansions and compressions are treated equivalently by the geometric mean, unlike the arithmetic one.

The definition of the geometric mean in \eqref{geomMeanDefn} can be shown to satisfy the Ando--Li--Mathias properties, which have been proposed as necessary for a reasonable definition of a geometric mean of positive-definite tensors \citep{Ando2004,Bhatia2012}. 
These properties, for the geometric mean of $\mathsfbi{C}:\Omega \rightarrow \Pos_3$, are listed below and provide mathematically rigorous constraints which lead to a well-behaved mean.
\begin{enumerate}
  \item Consistency with scalars in the commutative case. If $\mathsfbi{C}(x)\mathsfbi{C}(x')=\mathsfbi{C}(x')\mathsfbi{C}(x)$ for all $x,x' \in \Omega$ then 
  \begin{align}
  \gm{\mathsfbi{C}}=\exp \am{\log\mathsfbi{C}}.
  \end{align} 
  This property guarantees that (\ref{geomMeanDefn}) is a generalization of the standard scalar geometric mean.
  \item Joint homogeneity. If $f:\Omega \rightarrow \mathbb{R}_{>0}$ is a strictly positive scalar function on $\Omega$, then 
  \begin{align}
  \gm{f(x)\mathsfbi{C}(x)}=\underbrace{\left(\exp \am{\log f(x)} \right)}\gm{\mathsfbi{C}(x)}.
  \end{align}
  where the braced quantity is the scalar geometric mean of $f(x)$.
  \item Translation/scaling invariance.  For any $a,b \in \mathbb{R}$,
  \begin{align}
  \gm{\mathsfbi{C}(ax+b)}=\gm{\mathsfbi{C}(x)}.
  \end{align}
  The Ando--Li--Mathias properties are usually written for a discrete set of tensors. In that context, the translation/scaling invariance property is known as permutation invariance and refers to invariance of the mean under arbitrary re-ordering of the set. 
  \item Monotonicity. Given a positive-definite tensor-valued function $\mathsfbi{K}:\Omega \rightarrow \Pos_3$ such that $\mathsfbi{C}(x) \geq \mathsfbi{K}(x)$ for each $x \in \Omega$, then
  \begin{align}
  \gm{\mathsfbi{C}} \geq \gm{\mathsfbi{K}}.
  \end{align}
  Here and for the rest of the paper we assume the usual partial order (Loewner order) on the set of positive-definite tensors. 
  Monotonicity is especially helpful in ensuring that the normal components of the geometric mean do not take on unreasonable values: $ \gm{\mathsfbi{C}} \geq \gm{\mathsfbi{K}}$ guarantees that each of the normal components of $\gm{\mathsfbi{C}}$ is greater than or equal to the corresponding normal component of $\gm{\mathsfbi{K}}$\textemdash an intuitive requirement since $\mathsfbi{C}(x) \geq \mathsfbi{K}(x)$.
  \item Continuity. Given a sequence of positive-definite tensor-valued functions $\mathsfbi{C}_{(k)}:\Omega \rightarrow \Pos_3$ for $k\in \mathbb{N}$, such that $\lim_{k\rightarrow \infty} \mathsfbi{C}_{(k)}=\mathsfbi{C}$ in a pointwise sense, then
  \begin{align}
  \lim_{k\rightarrow \infty} \gm{\mathsfbi{C}_{(k)}}=\gm{\mathsfbi{C}}.
  \end{align}
  \item Joint concavity. Given a positive-definite tensor-valued function $\mathsfbi{K}:\Omega \rightarrow \Pos_3$, if $0 \leq \alpha \leq 1$
  \begin{align}
  \alpha \gm{\mathsfbi{C}} + (1-\alpha)\gm{\mathsfbi{K}} \leq \gm{\alpha \mathsfbi{C} + (1-\alpha)\mathsfbi{K}}
  \end{align}
  \item Congruence invariance\tz{,} or invariance under the action of the general linear group. For any invertible, constant $\mathsfbi{Y}$
  \begin{align}
  \gm{\mathsfbi{Y}\bcdot \mathsfbi{C} \bcdot \mathsfbi{Y}}^{\mathsf{T}} =
  \mathsfbi{Y}  \bcdot\gm{ \mathsfbi{C} }\bcdot \mathsfbi{Y}^{\mathsf{T}}. \label{congruence}
  \end{align}
  As mentioned previously, this is a physically appealing property which says that if $\mathsfbi{C}(x)$ is modified at each $x\in\Omega$ by a fixed deformation given by the deformation gradient $\mathsfbi{Y}$, then the geometric mean of the resulting modified conformation tensor is also modified similarly by $\mathsfbi{Y}$.
  \item Self-duality. The geometric mean satisfies the identity
  \begin{align}
  \gm{ \mathsfbi{C}^{-1} }^{-1}=\gm{ \mathsfbi{C} }
  \end{align}
  which implies that it treats compressions and expansions equivalently.
  \item Determinant identity. The determinant of $\mathsfbi{C}$ satisfies
  \begin{align}
  \det \gm{\mathsfbi{C}}=\exp \am{\log \det \mathsfbi{C}}  
  \end{align}
  where the right-hand side is the scalar geometric mean of $\det \mathsfbi{C}$.
  \item Arithmetic-geometric-harmonic means inequality. 
  \begin{align}
  \am{\mathsfbi{C}^{-1}}^{-1} \leq \gm{\mathsfbi{C}} \leq \am{\mathsfbi{C}}.
  \end{align}
\end{enumerate}
}


\citet{Moakher2005} obtained a more convenient expression for the geometric mean by explicitly solving the minimization problem \eqref{geomMeanDefn}.
It can be shown that $\gm{\mathsfbi{C}}$ satisfies
\begin{align}
\int_{\Omega} \log(\mathsfbi{C}^{-1}(x) \bcdot \gm{\mathsfbi{C}} ) \, \text{d}x = \mathsfbi{0} \quad
\Leftrightarrow &\quad
\int_{\Omega} \log(\mathsfbi{C}(x) \bcdot \gm{\mathsfbi{C}} ^{-1}) \, \text{d}x = \mathsfbi{0}. \label{Cgeomcond1}
\end{align}
Although \eqref{Cgeomcond1} is still  an implicit definition of $\gm{\mathsfbi{C}}$, it is useful in implementations.
Pre-multiplying the left equality in \eqref{Cgeomcond1} by $\gm{\mathsfbi{C}}^{1/2}$ and post-multiplying by $\gm{\mathsfbi{C}}^{-1/2}$ (and vice-versa for the right equality) yields
\begin{align} 
\int_{\Omega} \log( \mathsfbi{G}^{-1}(x)  ) \, \text{d}x = \mathsfbi{0}  \quad
\Leftrightarrow &\quad
\int_{\Omega} \log( \mathsfbi{G}(x)  ) \, \text{d}x = \mathsfbi{0} \label{Ggmcondition}
\end{align}
where we used the fact that $ \log (\mathsfbi{A}\bcdot \mathsfbi{B} \bcdot \mathsfbi{A}^{-1}) = \mathsfbi{A}\bcdot\log ( \mathsfbi{B}) \bcdot  \mathsfbi{A}^{-1}$ when $\mathsfbi{A}, \mathsfbi{B} \in \Pos_3$ and
\begin{align}
\mathsfbi{G}\equiv\gm{\mathsfbi{C}}^{-1/2}\bcdot\mathsfbi{C}\bcdot\gm{\mathsfbi{C}}^{-1/2}. \label{Ggm}
\end{align} 

The expression \eqref{Ggm} defines a fluctuating conformation tensor obtained by a geometric decomposition about $\gm{\mathsfbi{C}}$.
The geometric decomposition was recently introduced by \citet{Hameduddin2018a} as the appropriate definition of a fluctuating conformation tensor relative to a given nominal conformation. 
The condition \eqref{Ggmcondition} is a natural statistical characterization of such a fluctuating conformation tensor; it implies that $\am{\log  \mathsfbi{G}   }=\mathsfbi{0}$, or $\e{\am{\log  \mathsfbi{G} }}=\mathsfbi{I}$, which is the geometric analogue to the property that the arithmetic mean of (Euclidean) fluctuations in Reynolds decomposition is zero.
\rev{One of the Ando--Li--Mathias properties is a corollary: the volume of $\gm{\mathsfbi{C}}$ is equal to the scalar geometric mean of the volumes of $\mathsfbi{C}(x)$\textemdash an indication that the volume will be well-captured by $\gm{\mathsfbi{C}}$, unlike $\am{\mathsfbi{C}}$.}
Furthermore, since the volume is equal to the product of the principal stretches, the scalar geometric mean of the principal stretches of $\gm{\mathsfbi{C}}$ is equal to that of all principal stretches of $\mathsfbi{C}(x)$\textemdash an indication that the principal stretches of $\mathsfbi{C}(x)$ will also be well-captured by $\gm{\mathsfbi{C}}$, which suggests that $\tr\gm{\mathsfbi{C}}$ should be a good representative of $\tr {\mathsfbi{C}}$.
 

The geometric mean is an appealing quantity as it is obtained \rev{via an attractive geometry} and has useful properties.
However, there is no explicit way to calculate it and one must either solve a minimization problem \eqref{geomMeanDefn} or find the unique root of a multivariable nonlinear equation \eqref{Cgeomcond1}, both relatively more difficult than simple averaging.
\rev{An alternative to the geometric mean  which can be calculated explicitly is the log-Euclidean mean.
As with the geometric one, the log-Euclidean mean is also based on an underlying geodesically complete Riemannian structure on the set of positive-definite tensors \citep{Arsigny2006}. 
  The distance between positive-definite tensors $\mathsfbi{A}$ and $\mathsfbi{B}$ with this geometric structure is given by
  \begin{align}
  d_{\log}(\mathsfbi{A},\mathsfbi{B})\equiv \| \log \mathsfbi{A} - \log \mathsfbi{B} \|_{F}.   \label{geodesiclogDefn}
  \end{align}
  As with $d(\cdot,\cdot)$ defined in \eqref{geodesicDefn}, $d_{\log}(\cdot,\cdot)$ is invariant under inversions but it does not have the physically appealing property of invariance under the action of the general linear group. The resulting log-Euclidean mean can be calculated explicitly\textemdash typically also at a significantly reduced computational cost compared to the geometric mean\textemdash and is given by
\begin{align}
\lEm{\mathsfbi{C}} \equiv {\arg\min}_{\mathsfbi{W}\in\Pos_3} \int_{\Omega} d_{\log}^2(\mathsfbi{C}(x), \mathsfbi{W}) \, \text{d}x = \exp\left( \am{\log \mathsfbi{C}(x)}\right).
\end{align} 
}\rev{The log-Euclidean mean satisfies all of the Ando--Li--Mathias properties listed previously, except for monotonicity and invariance under the action of the general linear group.
The former property is important in ensuring that the normal components of $\mathsfbi{C}$ do not develop spurious peaks, while the latter ensures that an arbitrary deformation of conformation tensors results in the same deformation of their mean.
Interestingly, the log-Euclidean mean can be shown to be monotone under the chaotic order: $\mathsfbi{A} \geq \mathsfbi{B}$ if $\log \mathsfbi{A} -\log \mathsfbi{B}$ is positive-semidefinite
\citep{Seo2013}.}

\rev{
  The trace of the log-Euclidean mean is always guaranteed to be greater than that of the geometric mean \citep{Arsigny2005Thesis}.
  Since the determinants of both are equal by the determinant identity (see the ninth Ando--Li--Mathias property), this implies that the log-Euclidean mean is less isotropic than the geometric one. 
  In this sense, if the log-Euclidean and geometric means differ, the latter will be closer to the identity and hence represent a smaller deformation.
  By the first Ando--Li--Mathias property, the two means coincide when the principal axes of $\mathsfbi{C}$ are aligned and thus any difference between the two means arises due to rotation of the principal axes. 
}

Finally, for completeness, we define a square-root mean,
\begin{align}
\sqrtm{\mathsfbi{C}} = (\am{\mathsfbi{C}(x)^{\frac{1}{2}}})^2.
\end{align}
This mean is motivated by recent work that proposes using $\mathsfbi{C}^{\frac{1}{2}}$  in numerical schemes and for theoretical analysis since it simplifies some of the challenges associated with maintaining positivity of $\mathsfbi{C}$ \citep{Wang2014,Nguyen2016}.
In \S \ref{sec:dns}, we calculate the various means defined in this section and compare them to the arithmetic one.

\section{Means in viscoelastic drag-reduced turbulent channel flow}
\label{sec:dns}

\begin{figure}
  \centering
  \includegraphics[scale=0.9]{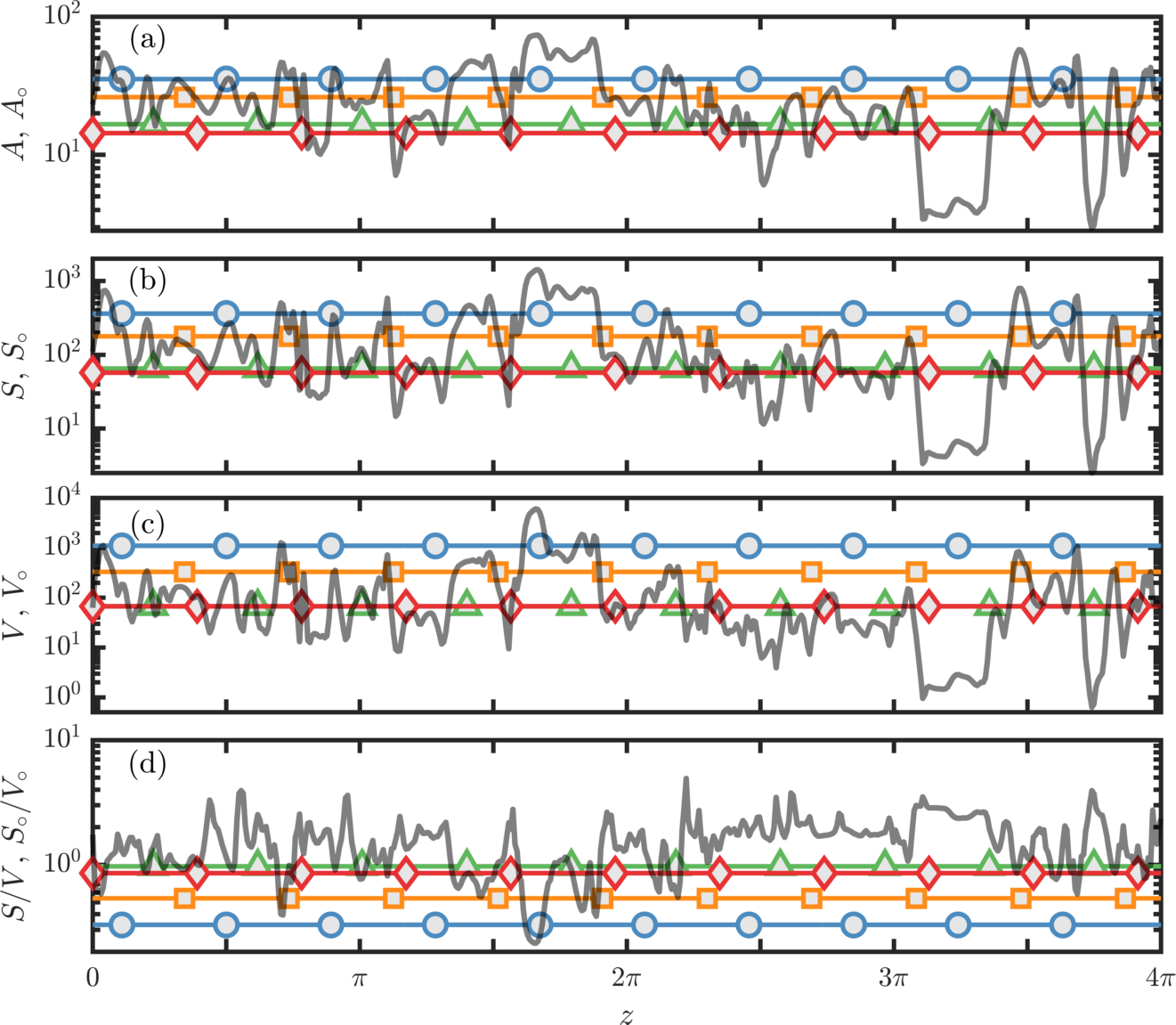}
  \caption{\rev{Characterization of conformation tensor: (a) sum of stretches, $A$ and $A_{\circ}$; (b) surface area, $S$ and $S_{\circ}$; (c) volume, $V$ and $V_{\circ}$; (d) surface area to volume ratio, $S/V$ and $S_{\circ}/V_{\circ}$. Here $\circ \in \{ \sum, \prod, \log, \sqrt{\,}\}$.  Fluctuating quantities along a $z$ traverse at $y^+=100$ (solid black line with 50\% transparency,{\color{gray}\linesolidsthick}); quantities calculated from various means along traverse: arithmetic 
  (circle markers, \linesolidsCirc
  ), geometric (diamond markers, \linesolidsDiam
  ), log-Euclidean (triangle markers, \linesolidsTri
  ), and square-root  (square markers, \linesolidsSqr
  ).     }}
  \label{fig:all_log_new}
\end{figure}

\rev{Figure \ref{fig:all_log_new} reproduces the right column of figure \ref{fig:am_shape_lin_log} and additionally shows the same quantities calculated from the alternative means outlined in \S\ref{sec:altmeans}, i.e.\,$A_{\circ}=\tr \sqrt{\circm{\mathsfbi{C}}}$, $V_{\circ}=  \sqrt{\det\circm{\mathsfbi{C}}}$, $S_{\circ}= (1/2)[(\tr \sqrt{\circm{\mathsfbi{C}}})^2 - \tr \circm{\mathsfbi{C}}]$ and $S_{\circ}/V_{\circ}$, where $\circ \in \{ \sum, \prod, \log, \sqrt{\,}\}$.}
While the arithmetic, log-Euclidean and square-root means were calculated using their explicit formulas, the geometric mean was calculated via \eqref{Ggmcondition} by numerical root-finding using the {\textsc{Matlab}} nonlinear least-squares solver with the trust-region-reflective algorithm. 
The search space was restricted to \rev{the feasible set $\Pos_3$ by re-formulating the optimization problem in terms of the logarithm of the geometric mean.}
The log-Euclidean mean was used to scale the search space and initialize the solver.

\rev{In figure \ref{fig:all_log_new}, quantities computed using the geometric and log-Euclidean means are all very similar.}
The stretch $A$, surface area $S$, and volume $V$ all appear to be over-predicted by the arithmetic mean, as was seen earlier, while the geometric and log-Euclidean means provide the smallest predictions of all three quantities.
The square-root mean lies in the middle and appears to capture $A$ relatively well, but not necessarily the other quantities.
Overall for these quantities, the alternative means appear to provide more representative approximation than the arithmetic mean, especially for $S$ and $V$, which are best captured by the geometric and log-Euclidean means. 
Both the arithmetic and square-root means show excessive sphericity (low $S/V$), while the geometric and log-Euclidean means are the most representative of this quantity.
\rev{These results are consistent with mathematical considerations that suggest that the geometric and log-Euclidean means are the most appropriate representative conformation tensors.}

In order to examine the behaviour of the fluctuations about various choices of the mean in a more quantitative fashion, we plot the joint probability density function (JPDF) of the logarithmic volume ratio (LVR) and the squared geodesic distance \rev{(c.f.\,equation \eqref{geodesicDefn})} of $\mathsfbi{C}$ from the chosen mean.
The LVR, $(1/2) \log (\det \mathsfbi{C}/\det \mathsfbi{\overline{C}})$, is the logarithm of the ratio of the volume of $\mathsfbi{C}$ to that of the chosen mean $\mathsfbi{\overline{C}}$.
The LVR is the worst predicted independent quantity by the arithmetic mean as suggested by figure \ref{fig:am_shape_lin_log}; this observation was found to hold more generally throughout turbulent channel flow by \citet{Hameduddin2018a}.
Those authors analysed the LVR because it naturally appears when the geometric structure of $\Pos_3$ is leveraged to analyse turbulent fluctuations of the conformation tensor. They found that the LVR with respect to the arithmetic mean was highly skewed to negative fluctuations.
In contrast, we expect the log-Euclidean and geometric means to much better represent the fluctuating volume as their determinants are guaranteed to be equal to the scalar geometric mean of $\det\mathsfbi{C}$.

\begin{figure}
  \centering
  \includegraphics[scale=0.95,trim={0 0 0 0},clip]{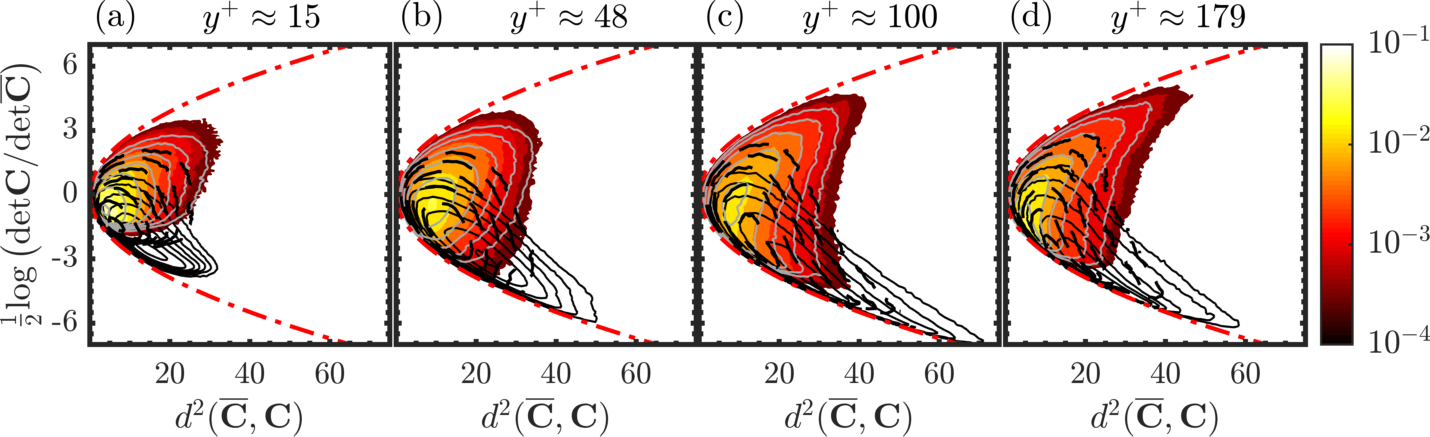}
  \caption{Joint PDF of LVR and squared geodesic for $\overline{\mathsfbi{C}}\in\{\am{\mathsfbi{C}},\gm{\mathsfbi{C}},\lEm{\mathsfbi{C}},\sqrtm{\mathsfbi{C}}\}$ at various wall-normal locations: (a) $y^+\approx 15$, (b) $y^+\approx 48$, (c) $y^+ \approx 100$, and (d) $y^+\approx 179$. Solid black contours (\linesolids) are  for $\am{\mathsfbi{C}}$, flood contours are for $\gm{\mathsfbi{C}}$, dashed black contours (\linedashed) are for $\sqrtm{\mathsfbi{C}}$, and solid grey contours ({\color{gray}\linesolids}) are for $\lEm{\mathsfbi{C}}$.  The red dashed dot lines ({\color{red}\linedshdot}) are realizability bounds that were derived by \citet{Hameduddin2018a}. }
  \label{fig:jpdf1_spacetime_new}
\end{figure}
 
The JPDF of LVR and squared geodesic distance is shown in figure \ref{fig:jpdf1_spacetime_new} at four wall-normal locations \rev{in the bottom half of the channel}. 
The JPDF and associated means were calculated using 12 snapshots spaced 10 convective time units apart from each other and by exploiting statistical homogeneity in the streamwise, spanwise, and temporal directions.
Reflection symmetry in $y$ was used to assess statistical convergence. 
\rev{The means calculated in one half of the channel were qualitatively similar to the other half, and quantitative differences were less than $12\%$ for any of the components. In all subsequent results, we only show results for the bottom half of the channel for maximum clarity.  
Some statistics, such as the mean velocity and certain components of the conformation tensor, converge much faster than others. We compared the arithmetic mean conformation tensor obtained from 12 snapshots spaced 10 convective time units apart with an average over all the simulation timesteps; all the components of the two arithmetic mean conformation tensors were qualitatively similar. Quantitatively, the $xx$ and $xy$ components matched within $10\%$;  the $yy$ and $zz$ components showed greater variations, up to $30\%$ for the former and $20\%$ for the latter.
These results are consistent with finite-time statistics of turbulent viscoelastic flows having low-frequency variations that can span very long time scales ($\mathcal{O}(10^3)$--$\mathcal{O}(10^4)$ convective time units).
}

The JPDF \rev{in figure \ref{fig:jpdf1_spacetime_new}} show that fluctuations with respect to the arithmetic mean are largely compressive with dramatic asymmetry about the zero LVR line, as reported by \citet{Hameduddin2018a}.
The square-root mean shows improvement vis-\`{a}-vis the arithmetic mean, but fluctuations remain excessively compressive.
The isocontours of fluctuations with respect to the geometric and log-Euclidean means are very similar to each other, and both are much more symmetric than those associated with the arithmetic and square-root means.
In particular, the problem with excessive compression disappears.
In addition, large excursions\textemdash defined as large geodesic deviations\textemdash are significantly reduced; the geodesic deviation for the most likely fluctuation is several times smaller for fluctuations defined with respect to the geometric and log-Euclidean means.
These properties had been elusive to date, because they are not satisfied by the arithmetic and square-root means, and are precisely captured by the geometric and log-Euclidean mean conformation tensors. 

The geometric and log-Euclidean means expose interesting behaviour: the most likely fluctuation in figure \ref{fig:jpdf1_spacetime_new} is volumetrically compressive but large excursions are most likely stretches, as evidenced by the isocontours tending towards the top right corner in each panel.
The latter tendency is minimized at $y^+\approx100$, where the isocontours are most symmetric about the zero LVR line, and where also the arithmetic and square-root means show the most compression.

\begin{figure}
  \centering
  \includegraphics[scale=1.0,trim={0 0 0 0},clip]{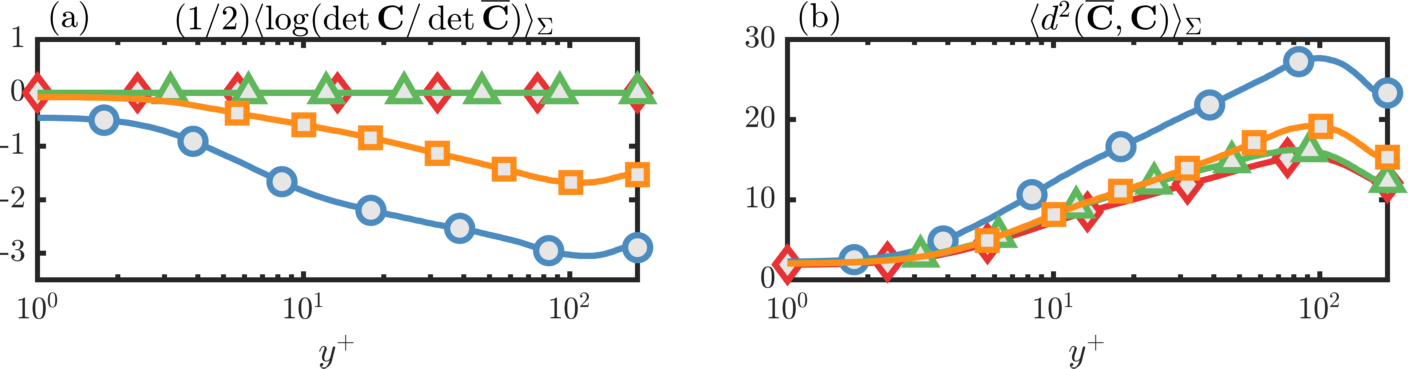}
  \caption{(a) Mean LVR, $(1/2)\am{\log (\det \mathsfbi{C}/\det \mathsfbi{\overline{C}})}$ and (b) mean squared geodesic distance from the mean, $\am{d^2(\mathsfbi{\overline{C}},\mathsfbi{C})}$, where $\mathsfbi{\overline{C}} \in \{ \am{\mathsfbi{C}}, \gm{\mathsfbi{C}}, \lEm{\mathsfbi{C}}, \sqrtm{\mathsfbi{C}} \}$.  See figure \ref{fig:all_log_new} for linetypes.      }
  \label{fig:G_scalars_two}
\end{figure}

Figure \ref{fig:G_scalars_two} shows the LVR and squared geodesic distance, averaged over the 12 snapshots and in the spanwise and streamwise directions, as a function of $y$ \rev{over the bottom half of the channel}.
As expected from figure \ref{fig:jpdf1_spacetime_new}, the average LVR of fluctuations with respect to the arithmetic mean is negative throughout the channel with a negative peak at $y^+\approx 100$.
The LVR for the square-root mean also peaks at $y^+\approx 100$ but shows relatively less drastic compressions.
As proved in \S\ref{sec:altmeans}, the geometric and log-Euclidean mean have zero LVR and thus naturally capture the volume well.

In figure \ref{fig:G_scalars_two}(b), the average squared geodesic distances of $\mathsfbi{C}$ from various means all feature a peak at $y^+\approx 100$. 
Such a peak was first reported by \citet{Hameduddin2018a} for the arithmetic mean only, but the present results demonstrate that it is independent of the way fluctuations are constructed.
The peak is significantly higher for the arithmetic mean than the alternative proposed ones, demonstrating its ill-suitability as a representative tensor.
The remaining means are relatively close to each other, with the geometric mean outperforming all others.

\begin{figure}
  \centering
  \includegraphics[scale=1.0]{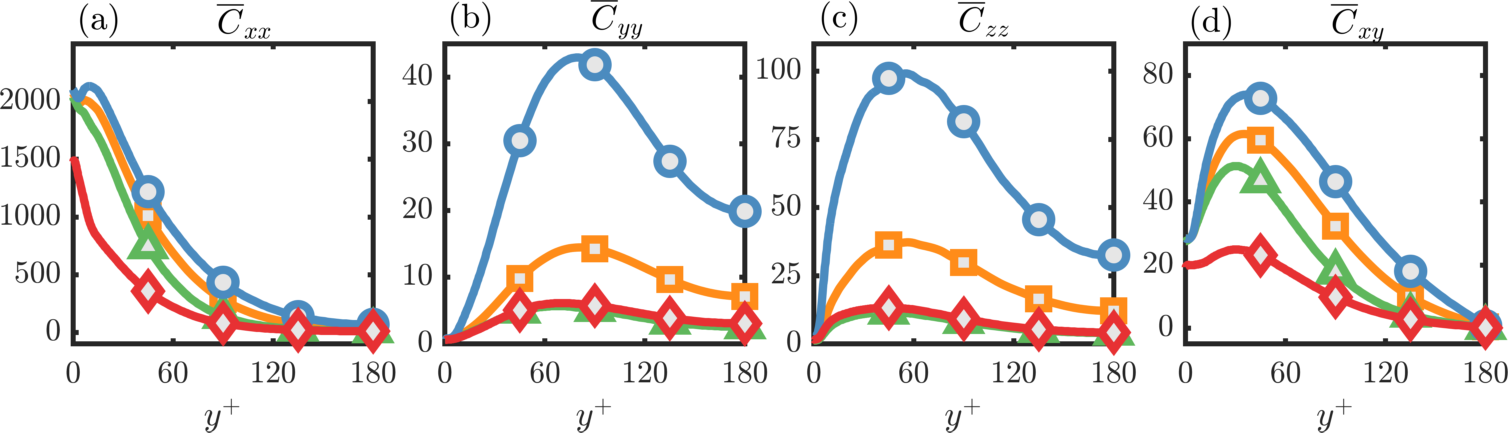}
  \caption{Components of various mean conformation tensors: (a) $\mathsfi{\overline{C}}_{xx}$, (b) $\mathsfi{\overline{C}}_{yy}$, (c) $\mathsfi{\overline{C}}_{zz}$, and (d) $\mathsfi{\overline{C}}_{xy}$, where $\mathsfbi{\overline{C}} \in \{ \am{\mathsfbi{C}}, \gm{\mathsfbi{C}}, \lEm{\mathsfbi{C}}, \sqrtm{\mathsfbi{C}} \}$.
    See figure \ref{fig:all_log_new} for linetypes.  }
  \label{fig:C_means}
\end{figure}

The non-zero components of the various mean conformation tensors are shown in figure \ref{fig:C_means}.
The components of $\am{\mathsfbi{C}}$ are the largest with the square-root mean showing the \rev{second largest} components.
The cross-stream stretches drop dramatically for the square-root mean, and even more for the geometric and log-Euclidean means, indicating that the arithmetic mean is an especially bad predictor of these stretches. 
The geometric and log-Euclidean mean are similar for $\mathsfi{C}_{yy}$ and $\mathsfi{C}_{zz}$ but show significant differences below $y^+\approx 100$ for $\mathsfi{C}_{xx}$ and $\mathsfi{C}_{xy}$.
\rev{Namely, both the $xx$ and $xy$ components are much larger for the log-Euclidean mean, with the latter component showing a peak at $y^+\approx 30$. 
As discussed earlier, if the log-Euclidean and geometric means do not coincide, the former is guaranteed to be more anisotropic and have a larger trace.
In this sense, it is unsurprising that the $xx$ component of the log-Euclidean mean seen in figure \ref{fig:C_means} is larger than that of the geometric mean.
}

\begin{figure}
  \centering
  \includegraphics[scale=1.0]{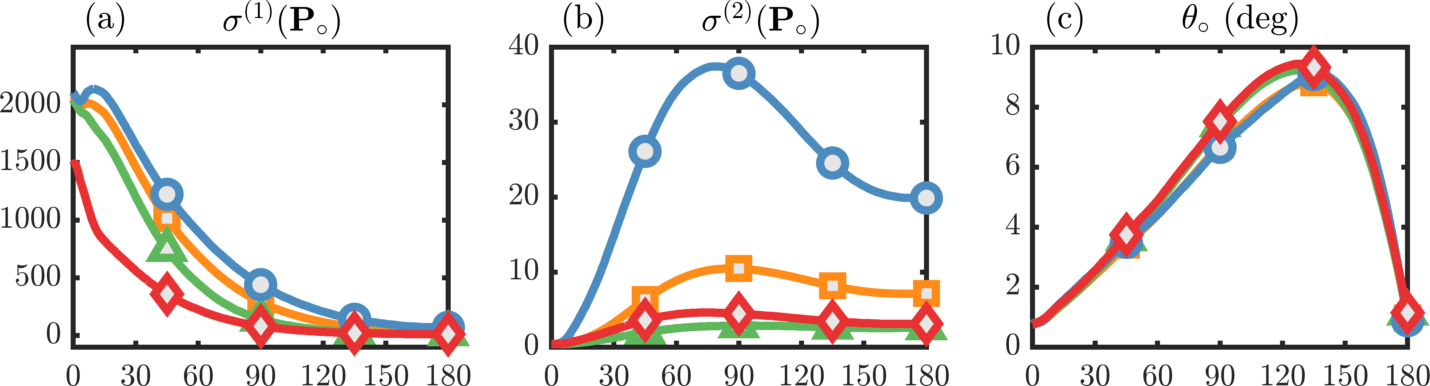}
  \caption{ \rev{(a) The largest eigenvalue of the tensor $\mathsfbi{P}_{\circ}$ as defined in \eqref{Pcircdefn}, (b) the second largest eigenvalue of $\mathsfbi{P}_{\circ}$, and (c) the rotation angle of the principal axes of $\mathsfbi{P}_{\circ}$ from the $x$ and $y$ axes. Here $\circ \in \{ \sum, \prod, \log, \sqrt{\,}\}$.
    See figure \ref{fig:all_log_new} for linetypes. } }
  \label{fig:C_means_eigtheta}
\end{figure}

\rev{
The peak at $y^+\approx 30$ in the $xy$ component of the log-Euclidean mean also appears in the geometric mean, but is much higher for the former. 
To identify the cause of its increase, we examine the principal axes and stretches of the mean conformation tensors.
Statistical symmetry requires that one of the principal directions is always aligned with the $z$ axis and, therefore, the $xz$ and $yz$ components of all the mean conformation tensors are guaranteed to be zero.
The principal stretch associated with this principal axis is simply the $zz$ component of the mean conformation tensor, which was already presented in figure \ref{fig:C_means}.
The remaining principal stretches and axes of $\circm{\mathsfbi{C}}$ can be identified with the aid of the reduced tensor $\mathsfbi{P}_{\circ}$ defined as
\begin{align}
\mathsfbi{P}_{\circ} \equiv
\begin{bmatrix}
\overline{\mathsfi{C}}_{xx} & \overline{\mathsfi{C}}_{xy} \\
\overline{\mathsfi{C}}_{xy} & \overline{\mathsfi{C}}_{yy} 
\end{bmatrix}, \quad \mathsfbi{\overline{C}}=\circm{\mathsfbi{C}}. \label{Pcircdefn}
\end{align}
Adopting the convention of ordering eigenvalues by magnitude, the remaining two principal stretches of $\circm{\mathsfbi{C}}$ are the eigenvalues of $\mathsfbi{P}_{\circ}$ and the respective principal axes of $\circm{\mathsfbi{C}}$ are the $x$ and $y$ axes rotated through an angle $\theta_{\circ}$.
It is then straightforward to obtain
\begin{align}
\mathsfi{\overline{C}}_{xy}= [\sigma^{(1)}(\mathsfbi{P}_{\circ})-\sigma^{(2)}(\mathsfbi{P}_{\circ})]\sin\theta_{\circ} \cos\theta_{\circ}  , \quad \mathsfbi{\overline{C}}=\circm{\mathsfbi{C}} \label{CxyPcirc}
\end{align}  
where $\sigma^{(k)}(\mathsfbi{P}_{\circ})$ denotes the $k$-th largest eigenvalues of $\mathsfbi{P}_{\circ}$.
It is clear from \eqref{CxyPcirc} that a large deviation of the log-Euclidean mean from the geometric mean in the stretch $\sigma^{(1)}(\mathsfbi{P}_{\circ})$, which is a good approximation for the $xx$ component of $\circm{\mathsfbi{C}}$ in viscoelastic turbulence, is sufficient to cause a discrepancy in $\mathsfi{\overline{C}}_{xy}$.
This insight is confirmed by figure \ref{fig:C_means_eigtheta} which shows $\sigma^{(k)}(\mathsfbi{P}_{\circ})$ for $k=1,2$ and $\theta_{\circ}$.
The stretch $\sigma^{(1)}(\mathsfbi{P}_{\circ})$ closely matches the $xx$ component shown in figure \ref{fig:C_means} for all the means, and thus shows a large difference between the log-Euclidean and geometric means.
At the same time, the $\sigma^{(2)}(\mathsfbi{P}_{\circ})$ and $\theta_{\circ}$ do not show as large a difference between the two means.
Thus, the larger peak in the $xy$ component of the log-Euclidean mean in figure \ref{fig:C_means} is largely a consequence of its greater $xx$ component. 
The discrepancy between the log-Euclidean and geometric means arises due to misalignment of the principal axes of $\mathsfbi{C}$ for different realizations, which primarily appears to exaggerate the largest stretch in the log-Euclidean mean.
}

Particularly interesting is the disappearance of the near-wall peak in $\mathsfi{C}_{xx}$ for the geometric and log-Euclidean means in figure \ref{fig:C_means}.
The near-wall peak in the arithmetic mean conformation tensor has been widely reported in the literature \citep[see, for instance, plots of the trace of $\am{\mathsfbi{C}}$ in][]{Housiadas2003} but our results suggest that it may be an artefact of the choice of the mean.
To understand how this peak and other broader differences between the arithmetic mean and the log-Euclidean (and geometric) means arise, we re-write the arithmetic mean in terms of the log-Euclidean mean.
\rev{Using the fact that $\log \mathsfbi{C} = \am{\log \mathsfbi{C}} +\bld{\mathcal{C}}'$, where $\bld{\mathcal{C}}'$ has a zero arithmetic mean, we have
\begin{align}
\am{\mathsfbi{C}}
 = \lEm{\mathsfbi{C}}
 +\am{\exp \bld{\mathcal{C}}'}  +
 \sum_{k=0}^{\infty} \mathsfbi{H}_{(k)} \label{amExpansion}
\end{align} 
where the higher-order terms are 
\begin{align}
\mathsfbi{H}_{(k)} \equiv \frac{1}{k!}\am{- (\am{\log \mathsfbi{C}}^{k} +   {\bld{\mathcal{C}}'}^{k}) + (\am{\log \mathsfbi{C}} + \bld{\mathcal{C}}')^{k}  }, \quad k=0,1,\hdots. 
\label{Hkdefn}
\end{align}
The first five terms of $\mathsfbi{H}_{(k)}$ are, 
\begin{align}
\mathsfbi{H}_{(0)} &= -\mathsfbi{I} \\
\mathsfbi{H}_{(1)} &= 0 \\
\mathsfbi{H}_{(2)} &= 0 \\
\mathsfbi{H}_{(3)} &= \frac{1}{6}\am{\bld{\mathcal{C}}'\bcdot \am{\log \mathsfbi{C}}\bcdot \bld{\mathcal{C}}'} + \frac{1}{3}\sym\left(\am{\log \mathsfbi{C}} \bcdot \am{{\bld{\mathcal{C}}'}^2} \right)    \\
\mathsfbi{H}_{(4)} &= 
\frac{1}{24} \left(  \am{\log \mathsfbi{C}} \bcdot \am{{\bld{\mathcal{C}}'}^2} \bcdot \am{\log \mathsfbi{C}} + 
  \am{\bld{\mathcal{C}}'\bcdot \am{\log \mathsfbi{C}}^2 \bcdot \bld{\mathcal{C}}'}
 \right)  \nonumber\\
&\hspace{1in} +\frac{1}{12}\sym \Big[
  \am{{\bld{\mathcal{C}}'}^3}\bcdot \am{\log \mathsfbi{C}}  + \am{\log \mathsfbi{C}}^2 \bcdot\am{{\bld{\mathcal{C}}'}^2}    \nonumber\\
&\hspace{2in} +
    \am{ 
    {\bld{\mathcal{C}}'}^2\bcdot \am{\log \mathsfbi{C}} \bcdot \bld{\mathcal{C}}' + (\am{\log \mathsfbi{C}} \bcdot \bld{\mathcal{C}}')^2  } \Big], 
\end{align}
where $\sym(\mathsfbi{A})= \frac{1}{2}(\mathsfbi{A} + \mathsfbi{A}^{\mathsf{T}})$ is the symmetric part of tensor $\mathsfbi{A}$.}

If one is to concede that the geometric mean is the most appropriate, and the log-Euclidean mean is an approximation, then \eqref{Hkdefn} conveys that the arithmetic mean is a function of the `correct' mean and the fluctuations about it. 
In other words, the arithmetic mean of the conformation tensor is a complex higher-order statistic when viewed from the appropriate statistical framework.
The wall-normal gradients of the arithmetic mean of the conformation tensor appear in the mean momentum equation directly, at least to first-order.
Their contribution can now be correctly interpreted as arising partially due to the direct effect of fluctuations.  
Thus, the arithmetic mean conformation tensor is similar in spirit to the Reynolds stress tensor, a higher-order statistic that is a direct function of the velocity fluctuations.

\begin{figure}
  \centering
  \includegraphics[scale=1.0]{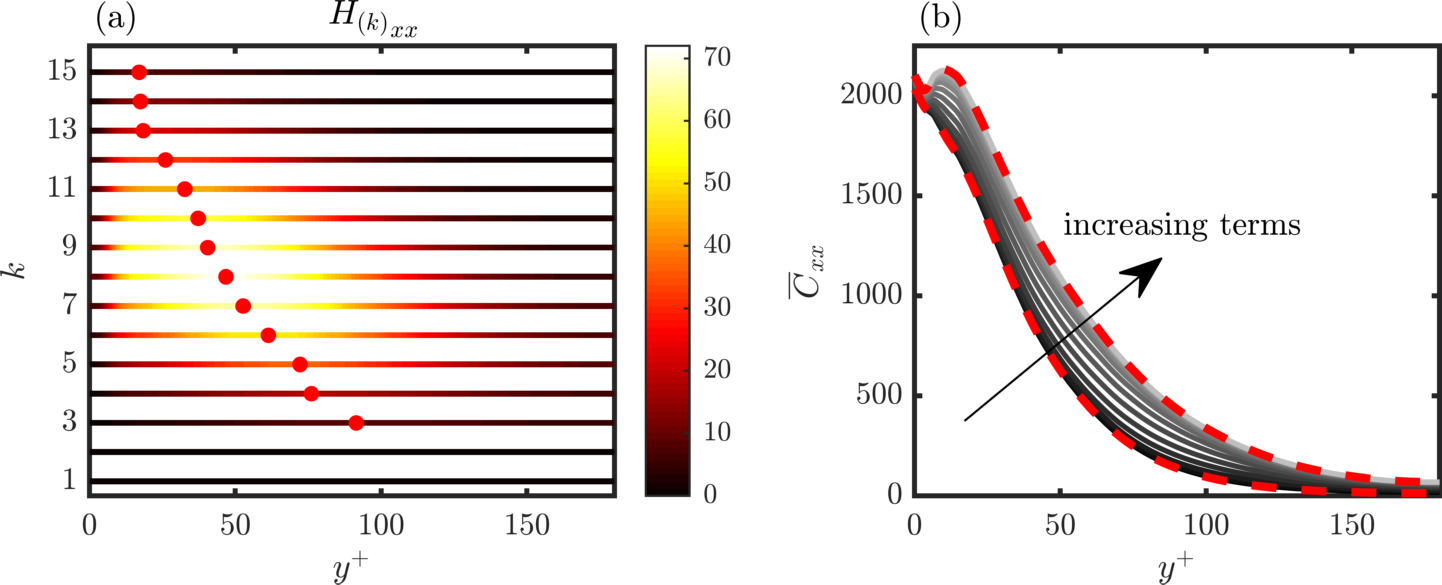}
  \caption{ \rev{(a) $xx$ component of $\mathsfbi{H}_{(k)}$ for $1 \leq k \leq 15$ with $\max \mathsfbi{H}_{(k)}$ marked with red filled circles ({\color{red}$\bullet$}), (b) $xx$ component of $\overline{\mathsfbi{C}}$ where $\lEm{\mathsfbi{C}}$ and $\am{\mathsfbi{C}}$ are the lower and upper red dashed lines ({\color{red}\linedashedthick}), respectively, and $\overline{\mathsfbi{C}} = \lEm{\mathsfbi{C}}
      +\am{\exp \bld{\mathcal{C}}'}  +
      \sum_{k=0}^{p} \mathsfbi{H}_{(k)}$ for $p=0,3,4,\hdots,14,15$ are the solid lines: dark to light solid lines ({\color{black}\linesolidsthick}  to {\color{gray}\linesolidsthick}) represent increasing terms from $p=0$ to $p=15$. }  }
  \label{fig:Hk_xx}
\end{figure}

\rev{
Figure \ref{fig:Hk_xx}(a) shows the streamwise ($xx$) component of $\mathsfbi{H}_{(k)}$ for $0 \leq k \leq 15$.
The maximum value taken on by ${\mathsfi{H}_{(k)}}_{xx}$ increases as $k$ increases up to $k=8$, after which it decreases.
At the same time, the location of the maximum moves closer to the wall. 
The cumulative effect of these higher-order terms is shown in figure \ref{fig:Hk_xx}(b), where the peak seen in the arithmetic mean emerges very clearly when all the terms up to  ${\mathsfi{H}_{(15)}}_{xx}$ are included in the expansion \eqref{amExpansion}.
Interestingly, the location of the maxima of ${\mathsfi{H}_{(k)}}_{xx}$ for $k\leq 15$ are always above $y^+= 17$ but the location of the peak in the arithmetic mean is at $y^+\approx 10$. 
This behaviour suggests that the peak in the arithmetic mean is not necessarily a consequence of increased polymer deformation localized at $y^+\approx 10$ but rather the cumulative effect of higher-order moments which have their own maxima elsewhere arising due to different reasons, e.g. rare events affect higher-order moments more.
}

\rev{
It is well-known that the existence of the near-wall peak in the $xx$ component of $\am{\mathsfbi{C}}$ is $\Wie$-dependent \citep{Housiadas2003}. 
In order to verify that the lack of this peak in the log-Euclidean and geometric means is not particular to the $\Wie=6.67$ considered so far, we also performed DNS of viscoelastic turbulent channel flow for $\Wie \in \{ 1.83,4.50,8.00 \}$.	
The algorithmic details are the same, as are all the flow parameters in table \ref{tab:SimParams} except for $\Wie$ and the resulting $\Rey_{\tau}$.
The initial conditions for the two cases $\Wie \in \{1.83,4.50\}$ were obtained from DNS of natural transition to turbulence via Tollmien--Schlichting waves reported by \citet{Lee2017}, and were extracted from the fully turbulent regime.  
In the case with $\Wie=8.00$, the initial condition was obtained by parameter continuation from $\Wie = 6.67$.  
In each simulation, the flow was further evolved for sufficiently long time using a positive-definiteness preserving algorithm \citep{Hameduddin2018a} in order to ensure that the turbulence is statistically stationary, prior to collecting statistics.
As with the $\Wie=6.67$ case, these statistics were calculated using $12$ snapshots, spaced $10$ convective time units apart from the last $130$ convective time units of each of the simulations described above, and by exploiting statistical homogeneity in the streamwise, spanwise, and temporal directions. 
Again, reflection symmetry in $y$ was used to assess statistical convergence. In addition, the arithmetic mean calculated from $12$ snapshots was compared with a continuous arithmetic average over all time-steps and minimal differences were found.
}

\begin{figure}
  \centering
  \includegraphics[scale=1.0]{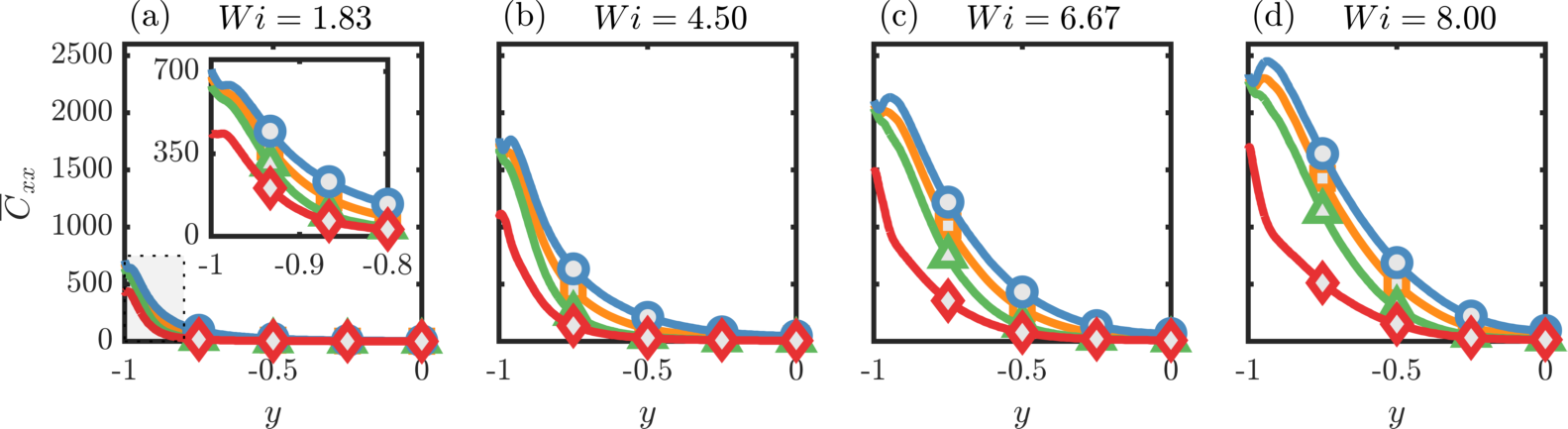}
  \caption{\rev{Streamwise component $\mathsfi{\overline{C}}_{xx}$ of the mean conformation tensor $\mathsfbi{\overline{C}}$ where $\mathsfbi{\overline{C}} \in \{ \am{\mathsfbi{C}}, \gm{\mathsfbi{C}}, \lEm{\mathsfbi{C}}, \sqrtm{\mathsfbi{C}} \}$ for varying $\Wie$. (a) $\Wie=1.83$, (b) $\Wie=4.50$, (c) $\Wie=6.67$, (d) $\Wie=8.00$. The inset in (a) is a close-up of the shaded area enclosed with dotted lines shown in (a). See figure \ref{fig:all_log_new} for linetypes. }} 
  \label{fig:Cxx_Wi_dependence}
\end{figure}

\rev{
Figure \ref{fig:Cxx_Wi_dependence} shows the $xx$ component of the various mean conformation tensors at different $\Wie$ for the bottom half of the channel only. 
The figure confirms the findings of previous authors that increasing $\Wie$ makes the near-wall peak of the $xx$ component of $\am{\mathsfbi{C}}$ more prominent; there is no peak away from the wall at $\Wie=1.83$, and $\Wie=8.00$ has the largest and most prominent peak. However, the $xx$ components of the log-Euclidean and geometric means both do not exhibit peaks away from the wall at any of the $\Wie$ in figure \ref{fig:Cxx_Wi_dependence}. Since $\Wie$ and associated levels of drag-reduction examined here are already considerably high, we predict that the $\Wie$-dependent phenomena that leads to the emergence of the peak in the arithmetic mean does not lead to a similar peak in the log-Euclidean and geometric means.  }  

\begin{figure}
	\centering
	\includegraphics[scale=1.0,trim={0 0 0 0},clip]{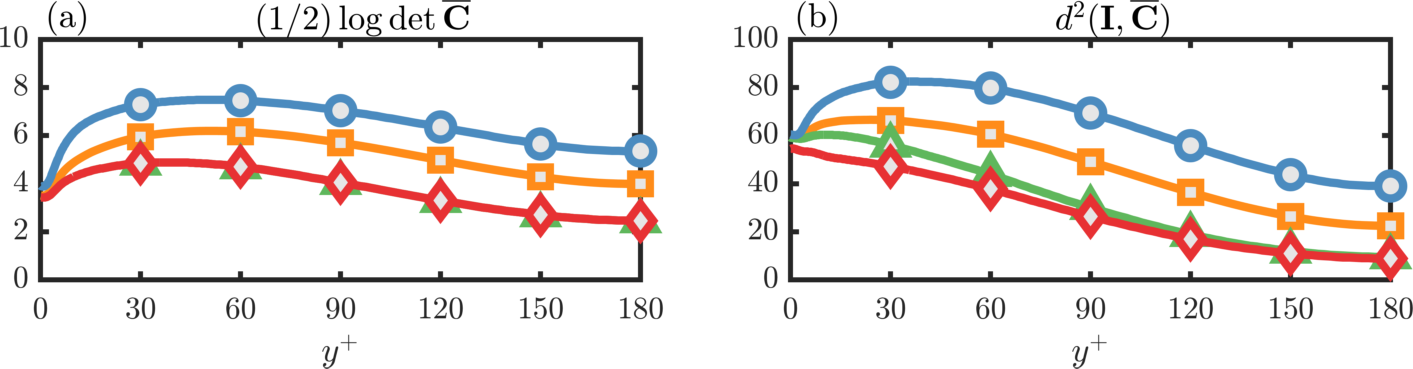}
    \caption{(a) Logarithmic volume, $(1/2)  \log \det \mathsfbi{\overline{C}}$ and (b) squared geodesic distance from the identity, $d^2(\mathsfbi{I},\mathsfbi{\overline{C}})$, where $\mathsfbi{\overline{C}} \in \{ \am{\mathsfbi{C}}, \gm{\mathsfbi{C}}, \lEm{\mathsfbi{C}}, \sqrtm{\mathsfbi{C}} \}$.  See figure \ref{fig:all_log_new} for linetypes.      } 
    \label{fig:C_scalars_two}
\end{figure}

The various mean conformation tensors \rev{at $\Wie=6.67$} are evaluated in more detail in figure \ref{fig:C_scalars_two}.
Here we show the logarithmic volume of the mean conformation tensor, $(1/2)  \log \det \mathsfbi{\overline{C}}$, and the squared geodesic distance from the identity, $d^2(\mathsfbi{I},\mathsfbi{\overline{C}})$.
As expected, the volume of the geometric and log-Euclidean means coincide.
This volume is several orders of magnitude smaller than that of the arithmetic mean, and is even smaller than the volume of the square-root mean.
Despite this small volume, which at the outset may appear to suggest that the resulting fluctuations must have large volume, recall that the LVR with respect to the geometric and log-Euclidean means was significantly smaller (in both a mean sense as well as in the probability density) compared to the LVR with respect to the arithmetic mean.
In this sense, the geometric and log-Euclidean means can very efficiently represent the volume of the conformation tensor.
A similar phenomenon can be seen in the squared geodesic distance from the identity.
Despite being relatively `small' in this sense, the fluctuations are geodesically closer to the geometric and log-Euclidean means as compared to the arithmetic and square-root means.

\section{Conclusions}
\label{sec:conclusions}

We demonstrated that the arithmetic mean is not representative of the ensemble of conformation tensors in viscoelastic turbulent flows and, therefore, we proposed alternative means\rev{, namely the geometric and log-Euclidean means.}
\rev{Both are based on a distinct well-behaved Riemannian structure on the set of positive-definite tensors \citep{Moakher2005,Arsigny2006}.} 
\rev{They were shown to possess useful properties enabling efficient representation of stretches and volume of the conformation tensor}\textemdash a key challenge with the arithmetic mean.
\rev{Additionally, the geometric mean satisfies the Ando--Li--Mathias properties that are considered important for any definition of a geometric mean of positive-definite tensors \citep{Ando2004} and also has the physically appealing property of being invariant under arbitrary, constant deformations of the conformation tensor.}
\rev{The log-Euclidean mean can also be viewed as a computationally efficient approximation of the geometric one}, and is of particular interest in light of recent efforts to develop numerical schemes that use the `log-conformation' approach \citep{Knechtges2015}.
The present work provides a geometric justification for analysis of turbulent flows using this approach.

We tested our proposed means in turbulent FENE-P channel flow and demonstrated that the geometric mean provides a more suitable alternative to the classical arithmetic mean.
It was not only geometrically the closest to the identity, but the fluctuations were also appreciably closer to it as compared to the arithmetic and square-root means. 
\rev{The log-Euclidean mean was also shown to be very similar to the geometric one in much of the channel.}

It has been widely reported in the literature on viscoelastic turbulent flows that a near-wall peak in the streamwise stretch of the arithmetic mean emerges and intensifies with elasticity.  
The geometric and log-Euclidean means did not show such behaviour, thus demonstrating that the apparent peak in polymer deformation is a consequence of the choice of the mean and not necessarily the increase in elasticity.  
By expressing the arithmetic mean in terms of the log-Euclidean mean and fluctuations about it, such differences between the means can be interpreted as the direct influence of higher-order statistical moments.
Namely, the arithmetic mean of the conformation tensor is not a physically intuitive fundamental quantity, but is rather a higher-order statistical quantity constructed via more physically fundamental elements.

The present work enables progress in turbulence modelling of viscoelastic flows by exposing the limitations of the arithmetic mean conformation tensor, and putting forward alternatives that are physically representative and more amenable to modelling.
The proposed alternative means can also be easily generalized to filters \citep{Bhatia2012}.
Such generalizations can be used to analyse and model subgrid stresses in large-eddy simulations of viscoelastic flows, and thus represent a potential opportunity for rapid advancement in computation of viscoelastic turbulence.

\bibliographystyle{jfm}
\bibliography{manuscript}

\begin{thebibliography}{17}
\expandafter\ifx\csname natexlab\endcsname\relax\def\natexlab#1{#1}\fi
\def\au#1{#1} \def\ed#1{#1} \def\yr#1{#1}\def\at#1{#1}\def\jt#1{\textit{#1}}
  \def\bt#1{#1}\def\bvol#1{\textbf{#1}} \def\vol#1{#1} \def\pg#1{#1}
  \def\publ#1{#1}\def\arxiv#1{#1}\def\org#1{#1}\def\st#1{\textit{#1}}

\bibitem[Ando {\em et~al.\/}(2004)Ando, Li \& Mathias]{Ando2004}
{\sc \au{Ando, T.}, \au{Li, C.} \& \au{Mathias, R.}} \yr{2004}  \at{{Geometric
  means}}.  \jt{Linear Algebra Appl.}  \bvol{385},  \pg{305--334}.

\bibitem[Arsigny {\em et~al.\/}(2005)Arsigny, Fillard, Pennec \&
  Ayache]{Arsigny2005Thesis}
{\sc \au{Arsigny, V.}, \au{Fillard, P.}, \au{Pennec, X.} \& \au{Ayache, N.}}
  \yr{2005}  \at{Fast and simple computations on tensors with log-{E}uclidean
  metrics.} PhD thesis, INRIA.

\bibitem[Arsigny {\em et~al.\/}(2006)Arsigny, Fillard, Pennec \&
  Ayache]{Arsigny2006}
{\sc \au{Arsigny, V.}, \au{Fillard, P.}, \au{Pennec, X.} \& \au{Ayache, N.}}
  \yr{2006}  \at{{Log-{E}uclidean metrics for fast and simple calculus on
  diffusion tensors}}.  \jt{Magn. Reson. Med.}  \bvol{56}~(2),  \pg{411--421}.

\bibitem[Arsigny {\em et~al.\/}(2007)Arsigny, Fillard, Pennec \&
  Ayache]{Arsigny2007}
{\sc \au{Arsigny, V.}, \au{Fillard, P.}, \au{Pennec, X.} \& \au{Ayache, N.}}
  \yr{2007}  \at{{Geometric means in a novel vector space structure on
  symmetric positive-definite matrices}}.  \jt{SIAM J. Matrix Anal. Appl.}
  \bvol{29}~(1),  \pg{328--347}.

\bibitem[Bhatia \& Karandikar(2012)]{Bhatia2012}
{\sc \au{Bhatia, R.} \& \au{Karandikar, R.~L.}} \yr{2012}  \at{{Monotonicity of
  the matrix geometric mean}}.  \jt{Math. Ann.}  \bvol{353}~(4),
  \pg{1453--1467}.

\bibitem[Hameduddin {\em et~al.\/}(2019)Hameduddin, Gayme \&
  Zaki]{Hameduddin2019a}
{\sc \au{Hameduddin, I.}, \au{Gayme, D.~F.} \& \au{Zaki, T.~A.}} \yr{2019}
  \at{Perturbative expansions of the conformation tensor in viscoelastic
  flows}.  \jt{J. Fluid Mech.}  \bvol{858},  \pg{377--406}.

\bibitem[Hameduddin {\em et~al.\/}(2018)Hameduddin, Meneveau, Zaki \&
  Gayme]{Hameduddin2018a}
{\sc \au{Hameduddin, I.}, \au{Meneveau, C.}, \au{Zaki, T.~A.} \& \au{Gayme,
  D.~F.}} \yr{2018}  \at{{Geometric decomposition of the conformation tensor in
  viscoelastic turbulence}}.  \jt{J. Fluid Mech.}  \bvol{842},  \pg{395--427}.

\bibitem[Hiai \& Petz(2009)]{Hiai2009}
{\sc \au{Hiai, F.} \& \au{Petz, D.}} \yr{2009}  \at{{{R}iemannian metrics on
  positive definite matrices related to means}}.  \jt{Linear Algebra Appl.}
  \bvol{430}~(11-12),  \pg{3105--3130}.

\bibitem[Housiadas \& Beris(2003)]{Housiadas2003}
{\sc \au{Housiadas, K.~D.} \& \au{Beris, A.}} \yr{2003}  \at{{Polymer-induced
  drag reduction: {E}ffects of the variations in elasticity and inertia in
  turbulent viscoelastic channel flow}}.  \jt{Phys. Fluids}  \bvol{15}~(8),
  \pg{2369--2384}.

\bibitem[Knechtges(2015)]{Knechtges2015}
{\sc \au{Knechtges, P.}} \yr{2015}  \at{{The fully-implicit log-conformation
  formulation and its application to three-dimensional flows}}.  \jt{J.
  Non-Newtonian Fluid Mech.}  \bvol{223},  \pg{209--220}.

\bibitem[Lang(2001)]{Lang2001}
{\sc \au{Lang, S.}} \yr{2001} {\em {Fundamentals of Differential Geometry}\/},
  \st{Graduate Texts in Mathematics},  \vol{vol. 191}.  \publ{New York, NY:
  Springer New York}.

\bibitem[Lee \& Zaki(2017)]{Lee2017}
{\sc \au{Lee, S.~J.} \& \au{Zaki, T.~A.}} \yr{2017}  \at{{Simulations of
  natural transition in viscoelastic channel flow}}.  \jt{J. Fluid Mech.}
  \bvol{820},  \pg{232--262}.

\bibitem[Masoudian {\em et~al.\/}(2013)Masoudian, Kim, Pinho \&
  Sureshkumar]{Masoudian2013}
{\sc \au{Masoudian, M.}, \au{Kim, K.}, \au{Pinho, F.~T.} \& \au{Sureshkumar,
  R.}} \yr{2013}  \at{{A viscoelastic {$k$--$\varepsilon$--$v^2$--$f$}
  turbulent flow model valid up to the maximum drag reduction limit}}.  \jt{J.
  Non-Newtonian Fluid Mech.}  \bvol{202},  \pg{99--111}.

\bibitem[Moakher(2005)]{Moakher2005}
{\sc \au{Moakher, M.}} \yr{2005}  \at{{A Differential Geometric Approach to the
  Geometric Mean of Symmetric Positive-Definite Matrices}}.  \jt{{SIAM} J.
  Matrix Anal. Appl.}  \bvol{26}~(3),  \pg{735--747}.

\bibitem[Nguyen {\em et~al.\/}(2016)Nguyen, Delache, Simo{\"{e}}ns, Bos \& {El
  Hajem}]{Nguyen2016}
{\sc \au{Nguyen, M.~Q.}, \au{Delache, A.}, \au{Simo{\"{e}}ns, S.}, \au{Bos, W.
  J.~T.} \& \au{{El Hajem}, M.}} \yr{2016}  \at{{Small scale dynamics of
  isotropic viscoelastic turbulence}}.  \jt{Phys. Rev. Fluids}  \bvol{1}~(8),
  \pg{083301}.

\bibitem[Seo(2013)]{Seo2013}
{\sc \au{Seo, Y.}} \yr{2013}  \at{Generalized {P}{\'o}lya--{S}zeg{\"o} type
  inequalities for some non-commutative geometric means}.  \jt{Linear Algebra
  Its Appl.}  \bvol{438}~(4),  \pg{1711--1726}.

\bibitem[Wang {\em et~al.\/}(2014)Wang, Graham, Hahn \& Xi]{Wang2014}
{\sc \au{Wang, S.}, \au{Graham, M.~D.}, \au{Hahn, F.~J.} \& \au{Xi, L.}}
  \yr{2014}  \at{{Time-series and extended {K}arhunen--{L}o{\`e}ve analysis of
  turbulent drag reduction in polymer solutions}}.  \jt{{AIChE}}
  \bvol{60}~(4),  \pg{1460--1475}.

\end{thebibliography}

%
%

\end{document}